\documentclass[epjCONF, columns]{svjour}
\usepackage{graphics}
\usepackage[varg]{txfonts} 
\usepackage[latin1]{inputenc}
\session-title{Conference Title, to be filled}
\begin{document}
\title{Performances of Anode-resistive Micromegas for HL-LHC}
\author{J. Manjarr\'es\inst{1}\fnmsep\thanks{\email{joany.manjarres-ramos@cea.fr}} \and T. Alexopoulos\inst{2} \and D. Atti\'e\inst{1} \and M.Boyer\inst{1} \and J. Derr\'e\inst{1} \and G. Fanourakis\inst{3} \and E. Ferrer-Ribas\inst{1} \and J. Gal\'an\inst{1} \and E. Gazis\inst{2} \and T. Geralis\inst{3} \and A. Giganon\inst{1} \and I. Giomataris\inst{1} \and S. Herlant\inst{1} \and F. Jeanneau\inst{1} \and Ph. Schune\inst{1} \and M. Titov\inst{1} \and G. Tsipolitis\inst{2} \and on behalf of MAMMA collaboration.}
\institute{ IRFU, CEA - Saclay, 91191 Gif-sur-Yvette Cedex, France \and National Tecnical University of Athens, Athens, Greece.\ \and Inst. of Nuclear Physics, NSCSR "Demokritos", Athens, Greece.}
\abstract{Micromegas technology is a promising candidate to replace Atlas forward muon chambers -tracking and trigger- for future HL-LHC upgrade of the experiment. The increase on background and pile-up event probability requires detector performances which are currently under studies in intensive RD activities.\\
We studied performances of four different resistive Micromegas detectors with different read-out strip pitches. These chambers were tested using $\sim$120 GeV momentum pions, at H6 CERN-SPS beam line in autumn 2010.  
For a strip pitch 500$\mu m$  we measure a resolution of $\sim$90 $\mu m$ and a efficiency of $\sim98\%$. The track angle effect on the efficiency was also studied. Our results show that resistive techniques induce no degradation on the efficiency or resolution, with respect to the standard Micromegas. In some configuration the resistive coating is able to reduce the discharge currents at least by a factor of 100.
} 
\maketitle
\section{Introduction}
\label{intro}
With the Large Hadron Collider (LHC) running and ramping in energy and luminosity, plans are already advancing for an upgrade. For the High Luminosity LHC (HL-LHC) project, the luminosity upgrade is expected to be increased  by a factor of ten \cite{sLHC}. 

To cope with the corresponding increase in background rates \cite{RTF}, Atlas experiment muon system will likely need major changes, at least in the highest rapidity region. Based on background estimations at HL-LHC, a list of requirements for these new detectors has been established:
\renewcommand{\labelitemi}{-}
\begin{itemize}
\addtolength{\itemsep}{-0.5mm}
  \item High counting rate capability, including dense ionization.
  \item High single plane detection efficiency, $\geqslant$ 98\%.
  \item Spatial resolution better than 100 $\mu$m, possibly up to large incident angles, 45$^{\circ}$.
  \end{itemize}

The MAMMA\footnote{Muon Atlas MicroMegas Activity} collaboration is focused on the development of large-area muon detectors based on bulk Micromegas\footnote{Micro-MEsh GASeous Detector}  technology as candidates for such an upgrade \cite{Micromegas1}.  In this paper, after a short introduction to the resistive Micromegas detector,  the analysis of the autumn 2010 test beam data is presented.




\section{Resistive-anodes Micromegas Detectors}
\label{sec:1}

In standard Micromegas detector, a very high amplification field is applied in a very thin gap \cite{Micromegas1}, as it is represented on the Fig. \ref{Detector}. In HL-LHC environment, the high flux of hadrons can produce highly ionizing particles that leads to large energy deposit and an increasing probability for sparks occurrence \cite{Raether}\cite{Sebas}.
Sparks are not a problem concerning the detector robustness or the electronics on which we can had a suitable protection, but the discharge of the whole micromesh induces a dead time up to 1 or 2 ms to recover the nominal voltage. One of the suggested solutions to reduce the effect of sparks is to use resistive coatings on top of the read-out strips. 
During this test we studied three different resistive-anodes and geometries configurations:


\begin{figure}[!htbp]
\vspace*{-7mm}
\begin{center}
\hspace*{8mm}
\resizebox{0.9\columnwidth}{!}{
\includegraphics{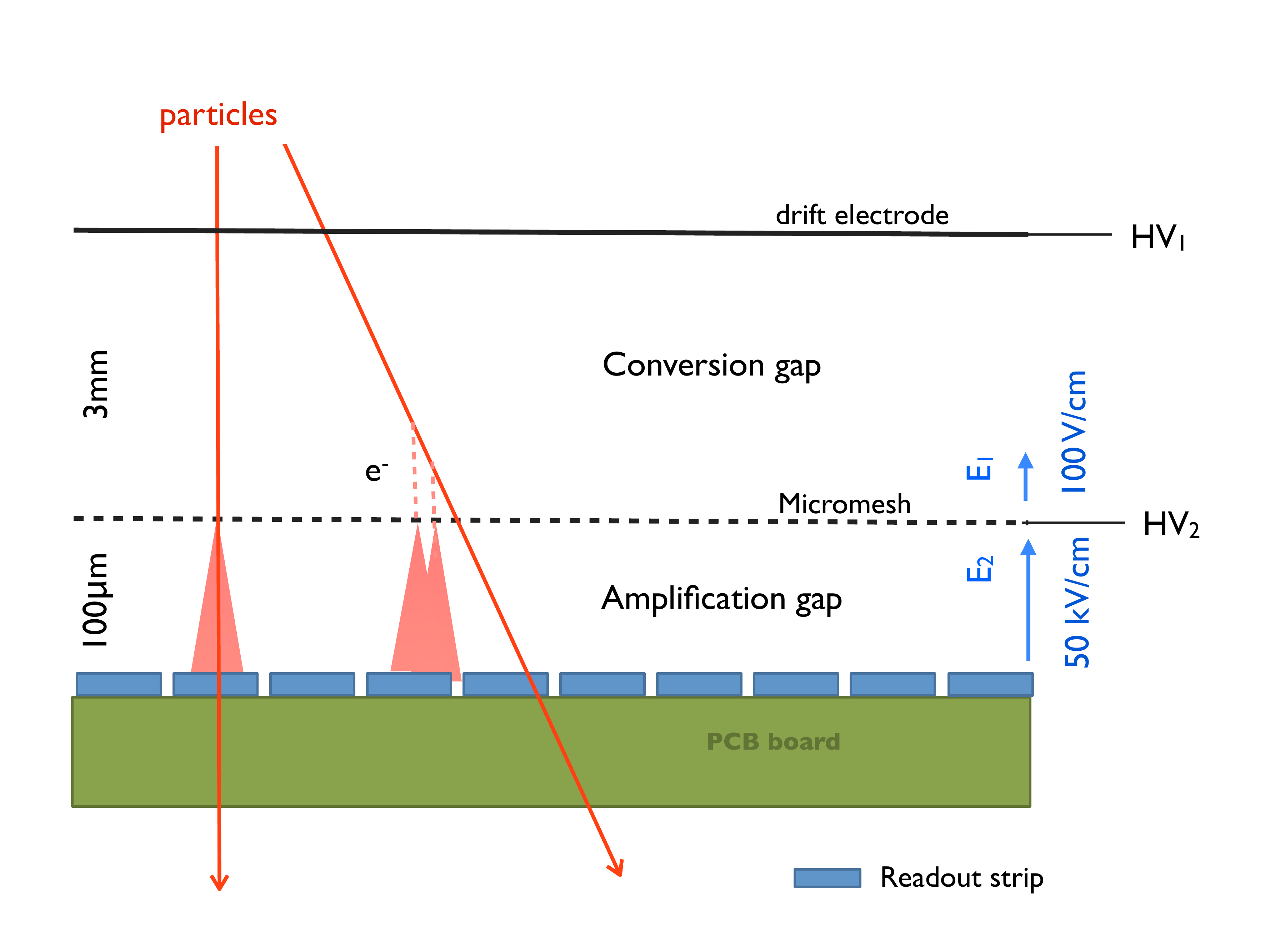}}
\caption{ Principle of a Micromegas detector. Sparks can occur between the anode and the micromesh. }
\vspace*{-3mm}
\label{Detector}
\end{center}
\end{figure}

\begin{figure}[!htbp]
\begin{center}
\vspace*{-3mm}
\resizebox{1.2\columnwidth}{!}{
\includegraphics{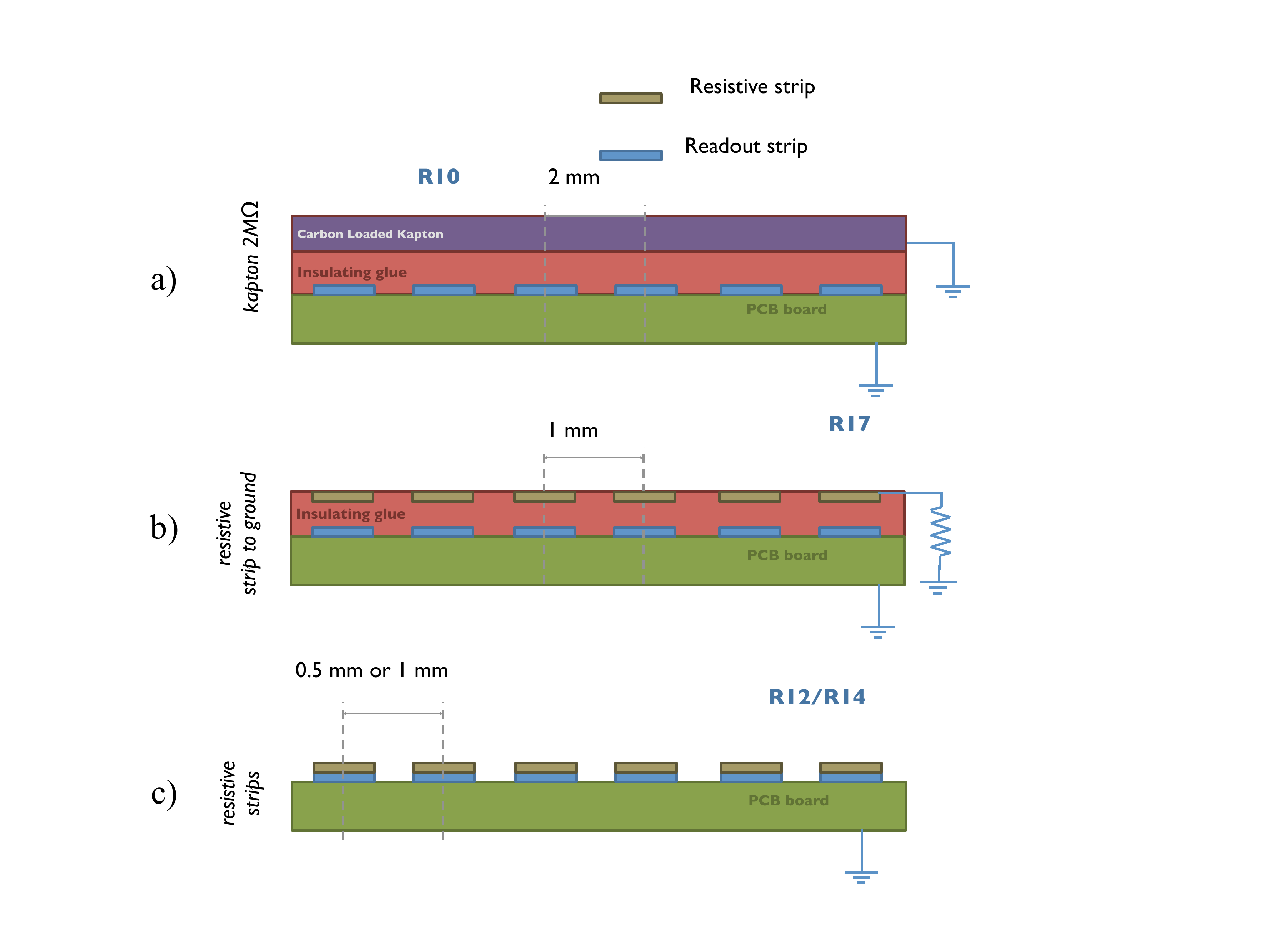}}
\vspace*{-5mm}
\caption{ Geometry of the different resistive-anode technologies.  }
\vspace*{-5mm}
\label{resist}
\end{center}
\end{figure}

\begin{itemize}

\item {\bf Kapton 2 $M\Omega / \Box$ layer:} In this case we have on top of the readout strips a insulating layer of 75$\mu$m thickness, on which is glued a foil of carbon-loaded kapton (2 $M\Omega / \Box$ resistive layer) see Fig. \ref{resist} $a$).

\item {\bf Resistive strip to ground:} The plane of strips is covered with an insulating layer of 64 $\mu$m thickness, on which resistive strips, matching the geometry of the copper strips, are deposited using a resistive ink of 100k$\Omega / \Box$. Each strip is grounded through a resistor of 30 M$\Omega$ and the resistor along the strips is 250 M$\Omega$ \cite{R_to_ground} see Fig. \ref{resist} $b)$.

\item {\bf Resistive strips:} On each copper-strip anode is deposited a resistive coating (resistive ink of 100k$\Omega / \Box$). The resistor along the strips is 300 k$\Omega$, see Figure \ref{resist} $c)$.

\end{itemize}
The specifications of the different technologies are summarized in the table \ref{Char_Res_MM}.

\begin{table}[!hbt]
\caption{Characteristics of the four resistive chambers tested.}
\begin{center}
\hspace*{-0.75cm}
\begin{tabular}{c p{1.cm} p{1.6cm}  c   c} 
\hline \\{ Chamber} & {pitch}  &{ Circuit type} & { Capacitance}  & { Gain max}  \\ 
\hline
 R10 & 2.0 mm &  kapton layer 2M$\Omega/ \Box$   & 1.67 nF & 7829 (410v)\\ 
 R17 & 1.0 mm & res. strip to ground & 943 pF & 10236 (410v)\\ 
 R14 & 1.0 mm & res. strips 300 k$\Omega/\Box$    & 943 pF & 10023 (410v)\\ 
R12 & 0.5 mm & res. strips 300 k$\Omega/ \Box$  & 637 pF  & 9835 (410v)\\ 
\hline 
\end{tabular}
\label{Char_Res_MM}
\end{center}
\vspace*{-0.95cm}
\end{table}

\section{The 2010 test beam}

The resistive chambers described above, were subsequently exposed to pions with momentum of $\sim$120 GeV, at CERN SPS H6 beam line, in autumn of 2010. An external reference measurement, on the plane normal to the beam direction was given by a telescope consisting of  3 (X-Y) plans of standard Micromegas. Four resistive detectors were tested in a two (X-Y). 
During the test beam, two different gas mixtures were used: Ar + 2\%C$_{4}$H$_{10}$+3\%CF$_{4}$ and  Ar + 2\%C$_{4}$H$_{10}$ for the telescope. The resistive detectors were tested with different high voltage values and mounted on a rotating structure, in order to collect data with different beam angle. The trigger was defined by a three scintillators coincidence, and the signal readout was performed with a GASSIPLEX electronics \cite{Gassiplex}. The data acquisition system was based on 4 CRAMS modules driven by a sequencer from CAEN.


\subsection{Performances}

The goal of this tests was to evaluate the performances of the resistive-anodes detectors and check the influence of resistive material on the spatial resolution and efficiency of the detector. The details of the data analysis can be found in \cite{yo}

\begin{figure}[!hbt] 
\begin{center}
\resizebox{0.9\columnwidth}{!}{
\includegraphics{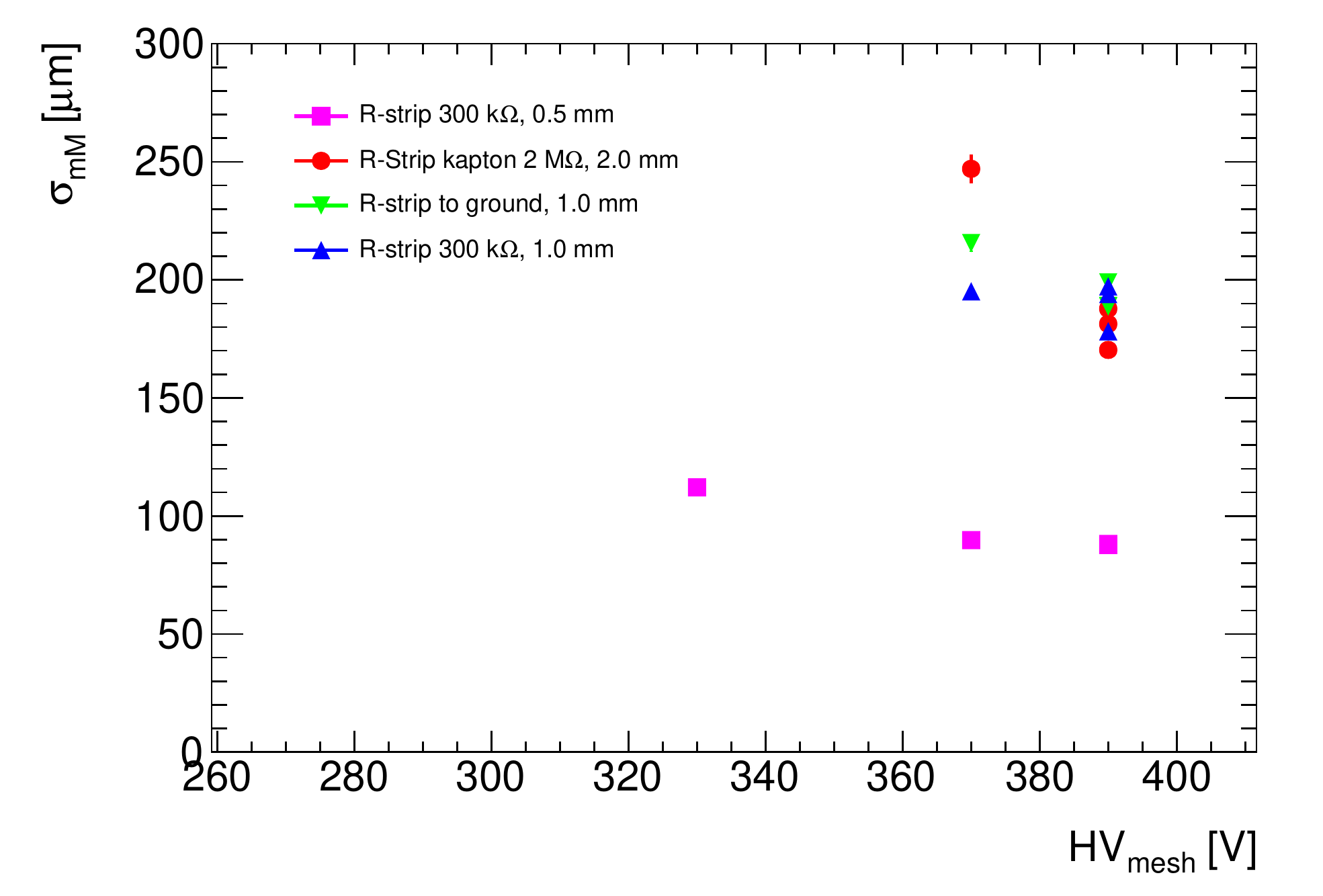} }
\caption{Spatial resolution vs the micro-mesh HV for three of the technologies tested.} 
\label{Residual_fit}
\end{center}
\vspace{-1.0cm}
\end{figure}

The spatial resolution is calculated using reference tracks given by the telescope. The results plotted in the Fig. \ref{Residual_fit} left are obtained by extrapolated these tracks at the level of the detector of interest. Then the distribution of the difference between the position measured and the position extrapolated is fitted by a gaussian curve of which the r.m.s. is the spatial resolution. This is clearly related to the pitch of the strips and the best resolution $\sigma_{Mm}$=(88.1 $\pm$ 0.7$\mu$m) is obtained for a pitch of 0.5 mm.

\subsection{Efficiency}

\begin{figure}[!hbt] 
\begin{center}
\resizebox{0.9\columnwidth}{!}{
\includegraphics{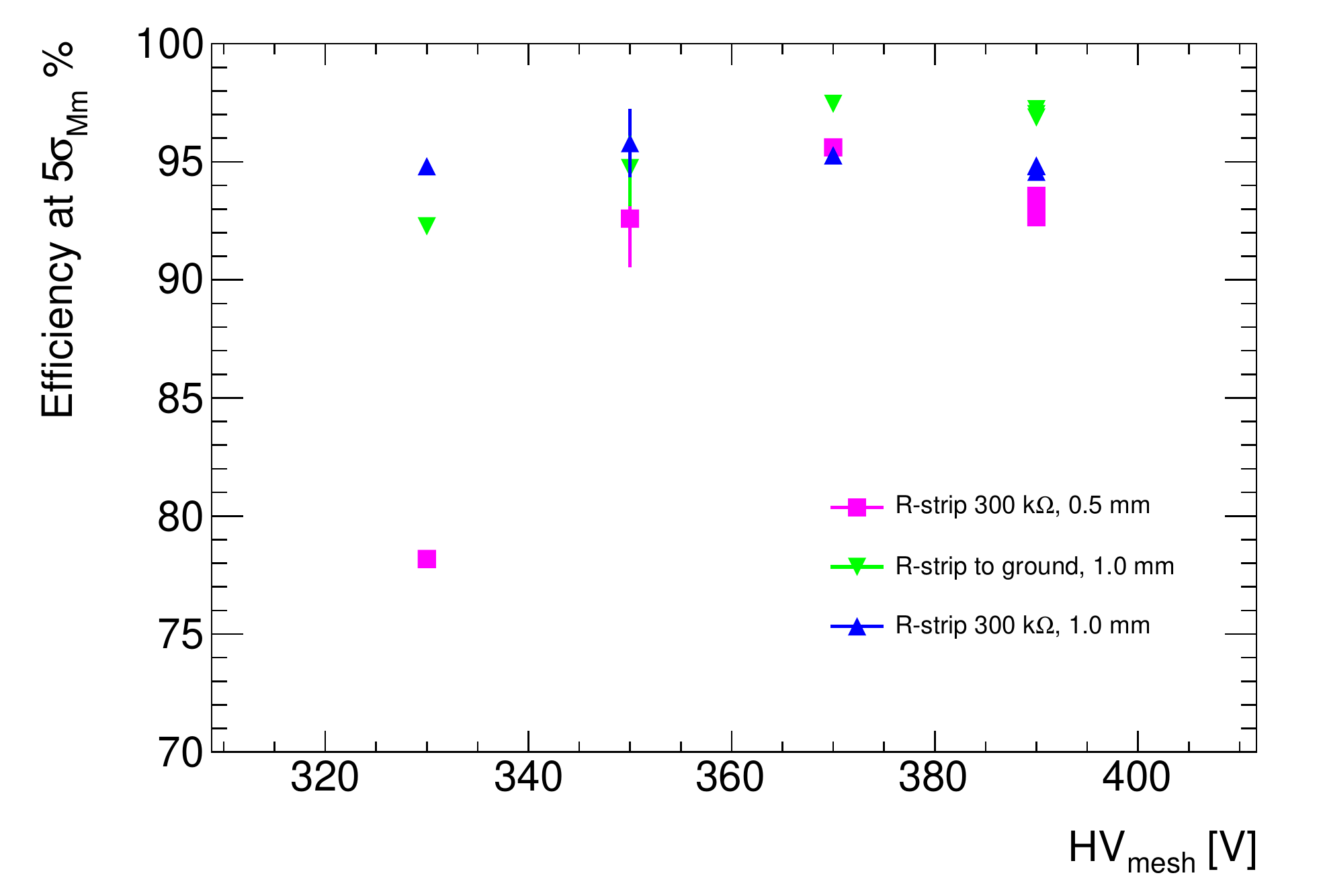}}
\caption{Efficiency $\sigma_{Mm}$ vs the micro-mesh HV  for un acceptance window of 5$\sigma_{Mm}$ for three of the technologies tested.} 
\label{eff}
\end{center}
\end{figure}

The detector is efficient when the measured position is within a window of $\pm$5$\sigma_{Mm}$ around the extrapolated position from the reference track of the telescope. As it is shown on Fig. \ref{eff} rigth, the detection efficiency increases with the mesh voltage up to 98\%. The detector with resistive strips connected to the ground (R-strip to the ground - R17) present the best efficiency results.

\begin{figure}[!hbt] 
\begin{center}
\resizebox{0.9\columnwidth}{!}{
\includegraphics{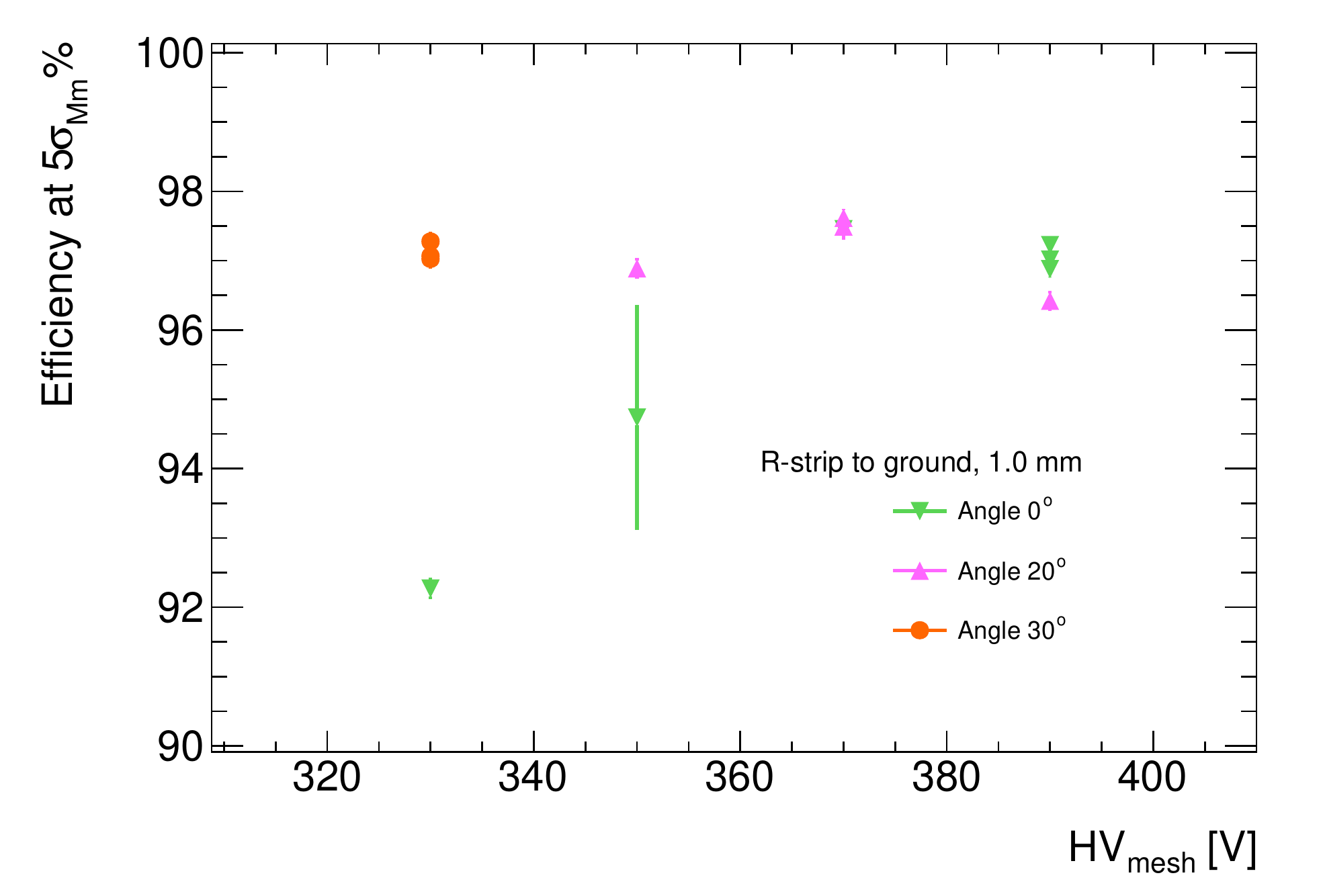}}
\caption{Efficiency vs micro-mesh voltages at several incidence angles for the R-strip to ground detector.} 
\label{eff_ang}
\end{center}
\vspace{-1.0cm}
\end{figure}

In order to get tracks with different incidence angles the detectors were rotated. As the signal is spread on several strips (depending on the pitch, see Fig. \ref{Detector}, we have to study the impact on the global efficiency. The Fig. \ref{eff_ang} shows the efficiency at different voltages on the micro-mesh and incidence track angle, for the R-strip to the ground detector (R17). For normal incidence tracks, the efficiency is around 98\%. For inclined tracks (the usual case in Atlas), the efficiency is of the same order and can be even better for low voltage values, because for inclined tracks the depth of gas is more important leading to more primary electron-ion pairs. Since this detector (R- to ground - R17) has the best efficiency, it is the best candidate to be proposed for the Atlas upgrade.

%
%

\subsection{Sparks behaviour}

\begin{figure}[!hbt]

\begin{center}
\hspace*{-.8cm}
\resizebox{1.2\columnwidth}{!}{
\includegraphics{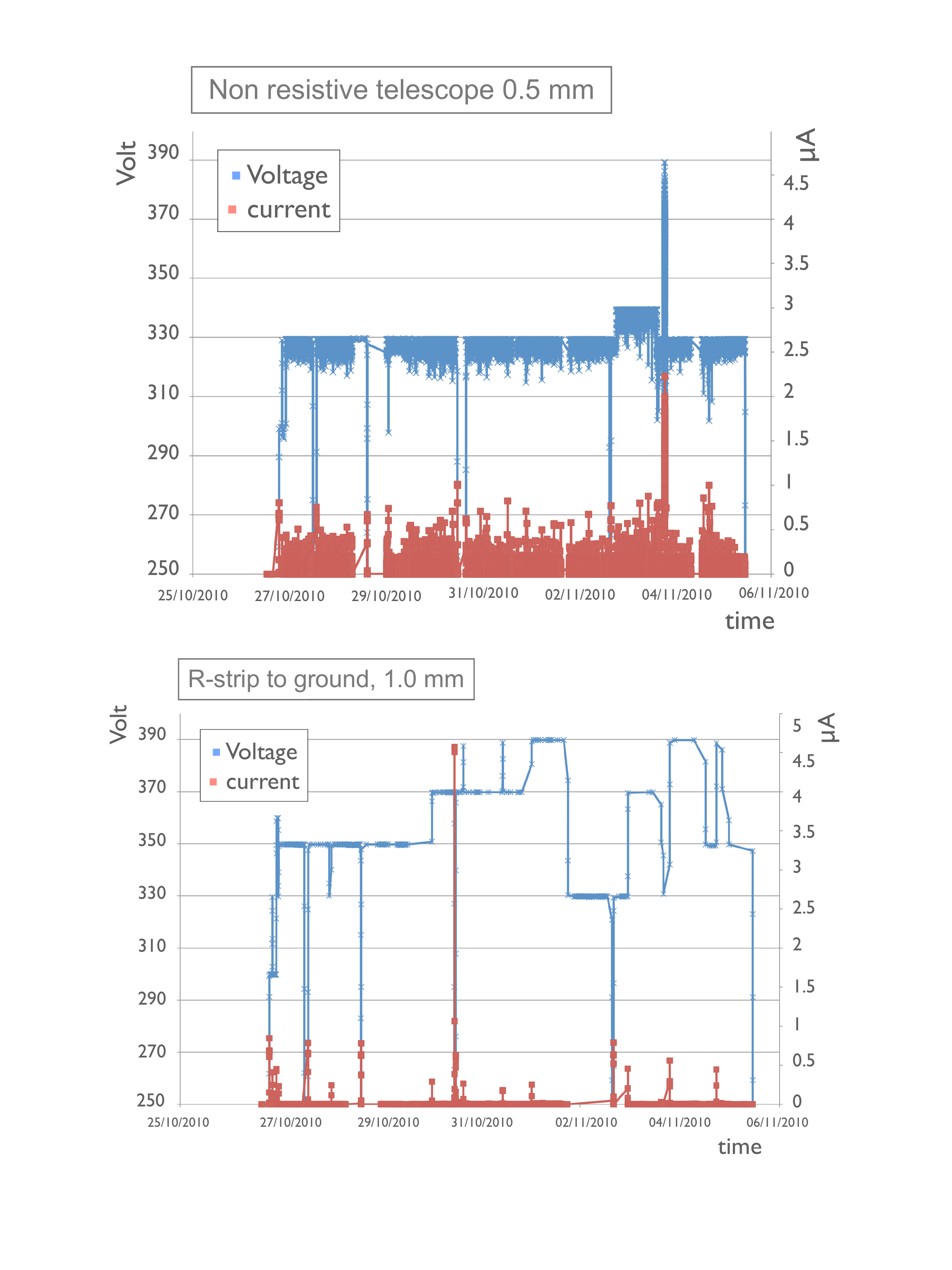}}
\caption{The monitored HV and currents as a function of the mesh HV during the test beam of autumn 2010 for a non-resistive Micromegas (top) and the chamber with resistive strips to ground R17 (bottom). The continuos line is show the HV, the points the current. }
\label{spark}
\end{center}
\hspace*{-.9cm}
\end{figure}

During the ten days of the beam test, the voltages and currents were monitored by the power supply (CAEN SY2527). The results for the voltage and current on the mesh are given on Fig. \ref{spark}. For a standard Micromegas detector -on the top - many current peaks and so many sparks are visible (red curve). The current can go up to 1 or 2 $\mu$A and are related to the voltage drops (blue curve) of 10 to 20 volts, that lead to a loss of efficiency of this detector. On the right, the same curves for the R-strip to ground detector (R17) show few sparks and very low current mainly corresponding to the charging up of the micro-mesh. For this detector there is no voltage drop thus no dead time and no loss of efficiency.

\section{Conclusion}

The different anode-resistive technologies that have been tested show very good performances, with a good spatial resolution (better than 100 $\mu$m for a pitch of 0.5 mm), a good efficiency (better than 95\%) and an efficient sparks reduction.
The R-strip-to-ground technology (R17) seems to be the best candidate for the HL-LHC environment since it presents a better efficiency, a very good spark reduction (no voltage drop - no dead time), a good spatial resolution achievable with a pitch of 0.5 mm, a better robustness (compared to R-coating) and no charging effect (compared to Carbon-Loaded-Kapton).

\end{document}